\documentclass[12pt]{article}
\usepackage{amsmath}
\usepackage{amsfonts}
\usepackage{amscd}
\usepackage{eucal}

\newcommand{\ddouble}{{\partial^{^{\kern-6pt \leftrightarrow}}}}

\newcommand{\beq}{\begin{equation}}
\newcommand{\eqn}[1]{\label{#1}\end{equation}}
\newcommand{\ba}{\begin{array}}
\newcommand{\ea}{\end{array}}

\def\psi{\Psi}

\def\inbar{\vrule height1.5ex width.4pt depth0pt}
\def\rlx{\relax\leavevmode}
\def\I{\leavevmode\hbox{\small1\kern-3.8pt\normalsize1}}
\def\openone{\leavevmode\hbox{\small1\kern-3.3pt\normalsize1}}
\def\Ione{\rlx{\rm 1\kern-2.7pt l}}
\def\Ik{\rlx{\rm I\kern-.18em k}}
\def\IC{\rlx\leavevmode
             \ifmmode\mathchoice
                    {\hbox{\kern.33em\inbar\kern-.3em{\rm C}}}
                    {\hbox{\kern.33em\inbar\kern-.3em{\rm C}}}
                    {\hbox{\kern.28em\sinbar\kern-.25em{\rm C}}}
                    {\hbox{\kern.25em\ssinbar\kern-.22em{\rm C}}}
             \else{\hbox{\kern.3em\inbar\kern-.3em{\rm C}}}\fi}
\def\IP{\rlx{\rm I\kern-.18em P}}
\def\IR{\rlx{\rm I\kern-.18em R}}
\def\IN{\rlx{\rm I\kern-.20em N}}

\def\llsymbol#1{\@llsymbol{\@nameuse{c@#1}}}
\def\@llsymbol#1{\ifcase#1\or {}\or {'}\or {''}\or {'''}\or
   {''''}\or {'''''}\or  \else\@ctrerr\fi\relax}
\newcounter{contador}

\newcommand{\ol}\overline
\newcommand{\ti}\tilde
\newcommand{\wt}\widetilde
\newcommand{\wh}\widehat
\newcommand{\bv}\breve
\newcommand{\dg}\dagger

\newcommand{\C}{^{\mbox{\scriptsize c}}}

\newcommand{\be}{\begin{equation}}
\newcommand{\ee}{\end{equation}}
\newcommand{\bl}{\begin{eqnarray}&}
\newcommand{\el}{&\end{eqnarray}}
\newcommand{\bq}{\begin{eqnarray}}
\newcommand{\eq}{\end{eqnarray}}

\def\ptoday{{\ifcase\month 
\or January, \or February, \or March, \or April,\or May, 
\or June, \or July, \or August, \or September, \or October, 
\or November, \or December,\fi\ \number \year}}

\begin{document}

{\hfill
\parbox{40mm}{{ICEN-PS-00/06} \\ {\ptoday} \vspace{4mm}}

\begin{center}
{{\LARGE {\rm Scale Invariance in a Non-Abelian \\ [3mm]
Chern-Simons-Matter Model}}} 


\vspace{5mm}

{\large J.L. Acebal$\textsuperscript{(a)}$ and D.H.T. 
Franco$\textsuperscript{(b)}$} 

\vspace{0,5cm}

(a){\em Pontif\'{\i}cia Universidade Cat\'olica de Minas Gerais -- (PUC-MG) 
\\Departamento de Matem\'atica e Estat\'{\i}stica \\[0,5mm]Av. Dom Jos\'e 
Gaspar 500 - CEP:30535-610 - Belo Horizonte - MG - Brasil.} 

\vspace{0,5cm}

(b){\em Universidade Cat\'olica de Petr\'opolis -- (UCP) \\Grupo de 
F\'{\i}sica Te\'orica -- (GFT) \\[0,5mm]Rua Bar\~ao do Amazonas 124 - 
CEP:25685-070 - Petr\'opolis - RJ - Brasil.} 

\vspace{0,5cm} 

{\tt e-mail: daniel@gft.ucp.br; acebal@fisica.ufmg.br}

\end{center}

\begin{abstract}
The general method of reduction in the number of coupling parameters is 
applied to a Chern-Simons-matter model with several independent couplings. 
We claim that considering the asymptotic region, and expressing all 
dimensionless coupling parameters as functions of the Chern-Simons 
coupling, it is possible to show that all $\beta$-functions vanish to any 
order of the perturbative series. Therefore, the model is asymptotically 
scale invariant. 
\end{abstract}

\vspace{0,3cm}

\small{

keywords: Renormalization, Asymptotic invariance, Coupling reduction} 

\section{Introduction}

It is well-known that quantum field theories are deeply connected to the 
presence of infinities (apparently insurmountable). Although the 
renormalization program can overcome this problem in a mathematically 
proper way, there exists a general feeling to look for field models which 
are not plagued by infinities, and therefore allowing us the possibility of 
getting non-perturbative results more easily. A natural candidate field 
theory of interest are the so-called topological field theories. 

Topological field theories\footnote{See~\cite{bbr-pr91} for a general 
review and references.} are a class of gauge models interesting from a 
physical point of view. In particular, their observables are of topological 
nature. An important topological model that has received much attention in 
the last years is the Chern-Simons model in three dimensions~\cite{s-btc}. 
In contrast to the usual Yang-Mills gauge theories, Chern-Simons theories, 
which include the case of three-dimensional gravity~\cite{witten-grav}, 
have some remarkable features. The main property of this class of models 
lies on a very interesting perturbative feature, namely, its ultraviolet 
finiteness~\cite{witten-grav,guadagnini}. A complete and rigorous proof of 
the latter property has been given in~\cite{blasi} for $D=1+2$ Chern-Simons 
theories in the Landau gauge, and for the Chern-Simons-Yang-Mills theory 
in~\cite{CFNP0}. In the more general case where these models are coupled to 
matter fields, the Chern-Simons coupling constant keeps unrenormalized, the 
corresponding $\beta$-function being trivial. This has been rigorously 
proved in~\cite{maggiore} for the Abelian Chern-Simons theory coupled to 
scalar matter fields. 

At the same time, for the non-Abelian theory coupled to general scalar and 
spinorial matter fields, an argument based on the assumed existence of an 
invariant regularization has been proposed by the authors of 
ref.~\cite{pronin} to demonstrate the triviality of the Chern-Simons 
coupling. In~\cite{CFNP}, a rigorous proof has been given for the vanishing 
of the Chern-Simons $\beta$-function in the presence of general scalar and 
spinorial matter, where one avoids the necessity of invoking a particular 
regularization procedure, by using the ``algebraic'' method of 
renormalization~\cite{pisor}, which only relies on general theorems of 
renormalization theory. The proof relies on the use of a local, 
non-integrated version of the Callan-Symanzik equation~\cite{CFNP, Bar}. 

The scale invariance of non-Abelian Chern-Simons minimally coupled to 
matter has already been discussed, at the two-loop level, in the paper of 
ref.~\cite{CSW}. It is the main goal of our present work to extend this 
result. Indeed, we include matter with minimal gauge coupling, but we also 
adjoin to the action matter-matter interaction terms that respect the 
power-counting renormalizability. With this rather complete action, we 
investigate the possibility that such a model be completely finite, i.e.  
its self-coupling parameters are not renormalized, too. The idea is to use 
the technique of reduction of coupling parameters~\cite{OSZ,Oe} to 
unify all couplings. The nonrenormalization of the matter self-couplings 
would then follow from the nonrenormalization of the Chern-Simons coupling. 
Therefore, we are able to show the all-order asymptotic scale invariance 
of the model, without resorting to laborious diagrammatical 
analysis. By asymptotic scale invariance to all orders, we mean the absence 
of Callan-Symanzik coupling constants renormalization, to every order of 
perturbation theory. Anomalous dimensions are however allowed to be 
different from zero. The latter, corresponding to field redefinitions, 
are physically trivial and hence vanish on the mass-shell~\cite{Pi}. 

The outline of this letter is as follows. In the following, we display the 
main purpose of this work: considering a non-Abelian Chern-Simons-matter 
model in the asymptotic region, using the basic construction of the 
reduction of coupling parameters~\cite{OSZ,Oe}, we shall show that 
expressing all dimensionless coupling parameters as functions of a single 
parameter, in our case the Chern-Simons coupling, then all 
$\beta$-functions vanish to any order of perturbation theory. Therefore, 
the model is asymptotically scale invariant. In section 3, we draw our 
conclusions. 

\section{Criterion for the Scale Invariance}

As mentioned in the Introduction, it has been showed in~\cite{CFNP} that 
for the model (in the Landau gauge) 
\begin{align}
\Sigma={\int }d^3x\,\,&\{ \kappa \,\varepsilon
^{\mu \nu \rho }(A_\mu ^a\partial_\nu A_\rho ^a+\frac 13f_{abc}A_\mu
^aA_\nu ^bA_\rho^c) + (\,i\,\bar{\Psi}_{j}\gamma ^\mu D_\mu 
\Psi_j+\frac 12 D_\mu \varphi_i^* D^\mu 
\varphi_i - {\cal V}(\varphi,\,\Psi)) \nonumber\\[3mm]
&+ (\partial_\mu b^a A^{a\mu} + 
\partial_\mu \bar{c}_a D^\mu c^a) + 
\sum_{\Phi=A_\mu
^a,\,c^a,\,\Psi_j,\,\varphi_i}\Phi^{\natural} s\Phi \}\,\,,
\label{g-inv-action1}
\end{align}
the Chern-Simons coupling keeps unrenormalized to all orders. Here, the 
gauge field $A_\mu^a(x)$ lies in the adjoint representation of the gauge 
group $SU(N)$, with Lie algebra $\left[ X_a,X_b\right] =if_{ab}{}^cX_c$. 
The scalar matter fields $\varphi _i(x)$ and the spinor matter fields 
$\Psi_j(x)$ are in the fundamental representation of $SU(N)$, the 
generators being represented by the matrices $T_a^{(\varphi )}$ and 
$T_a^{(\Psi)}$, respectively. $c^a$, $\bar{c}^a$ and $b^a$ are the ghost, 
the antighost and the Lagrange multiplier fields, respectively. ${A}_{a 
\mu}^\natural$, ${c}_a^\natural$, ${\Psi}_j^\natural$, 
${\varphi}_i^\natural$ are the ``antifields'' coupled to the nonlinear 
variations under BRS transformations. The generalized covariant derivative 
is defined by 
\begin{equation}
D_\mu \Phi(x)=( \partial _\mu -i\,{A_\mu^a}(x)T_a^{(\Phi )})\Phi (x)\,\,. 
\end{equation}
The function ${\cal V}(\varphi ,\,\Psi )$ defines the self-interactions of
the matter fields and their masses: 
\begin{align}
{\cal V}\left(\varphi,\Psi \right)=&\frac 12\lambda_1\,\bar{\Psi}_j\Psi_j 
\varphi_k^* \varphi_k + \frac 12\lambda_2\,\bar{\Psi}_j\Psi_k 
\varphi_k^* \varphi_j + \frac 16\lambda_3\,(\varphi_i^*\varphi_i)^3
\nonumber \\[3mm]
&+\mbox{mass terms} + \mbox{dimensionful couplings}\,\,.  \label{potential} 
\end{align}
We refer to~\cite{CFNP} for more details. 

We are interested in showing that, under a particular circumstance, the 
model (\ref{g-inv-action1}) may be scale invariant. Scale invariance means 
here the  vanishing of the Callan-Symanzik $\beta$-functions. We retain the 
possibility of wave function renormalization, so that the anomalous 
dimensions are allowed to be different from zero. It means the scale 
invariance of physical quantities still holds since the anomalous 
dimensions, corresponding to field redefinitions, are physically trivial. 

Let us next outline the criterion for the model (\ref{g-inv-action1}) to be 
scale invariant. The starting point for our analysis is the Callan-Symanzik 
equation for the model (\ref{g-inv-action1})~\cite{CFNP}
\begin{equation}
\left( D + \beta_\kappa \partial_\kappa + \beta_{\lambda_1}\partial_{\lambda_1}
+\beta_{\lambda_2} \partial_{\lambda_2} + \beta_{\lambda_3} 
\partial_{\lambda_3} - \gamma_A{\cal N}_A-\gamma_\psi {\cal N}_\psi- 
\gamma_\varphi{\cal N}_\varphi \right) \Gamma\, \sim 0 \,,
\label{callan-symanzik}
\end{equation}
where
\[
D\Gamma=\sum\limits_{{\rm all\ dimensionful\ parameters\ \mu}} \mu 
\frac{\partial\Gamma} {\partial \,\mu }\,\,.
\]
The signal $\sim $ means equality up to mass terms and dimensionful 
couplings. Note that, for a moment, we are considering that $\beta_\kappa$ 
is not vanishing. 
 
In (\ref{g-inv-action1}) the independent dimensionless parameters are 
$\kappa, \lambda_1, \lambda_2, \lambda_3$. We call $\kappa$ the primary 
coupling. Demanding that the coupling constants $\lambda_1, 
\lambda_2, \lambda_3$ have to be functions of the coupling constant $\kappa$:
\begin{equation}
\lambda_r=\kappa f_r(\kappa) \quad r=1,2,3\,\,,
\end{equation}
where
\begin{equation}
f_r(\kappa)=f_r^{(0)}+\sum_{m=1}^\infty \chi_r^{(m)}\kappa^m\,\,,
\label{sol}
\end{equation}
is a formal power series solution (special solution of ref.\cite{Oe}), then 
any relation among couplings of this type can be expressed by reduction 
equations~\cite{OSZ,Oe} 
\begin{equation}
\beta_{\lambda_r}=\beta_\kappa \left(\kappa \frac{df_r}{d\kappa}
+f_r\right)\,\,.\label{requa}
\end{equation}
Since $\beta_\kappa$ is identically zero, then (\ref{requa}) implies that 
$\beta_{\lambda_r}=0$ to all orders of perturbation theory. The reduction 
equations (\ref{requa}), in general, are necessary and sufficient 
conditions insuring that the renormalizable model (\ref{g-inv-action1}), with 
several independent couplings, can be reduced to a renormalizable model 
where all couplings are functions which depend only upon the Chern-Simons  
coupling. 

For the sake of simplicity, we shall consider the action 
(\ref{g-inv-action1}) based on group the $SU(2)$. We get the following 
$\beta$-functions~\cite{AKK}:
\begin{align}
\beta_\kappa&=0\,\,,\nonumber\\[3mm]
\beta_{\lambda_1}&=\frac 9{16}\kappa^3-\frac 9{4}\lambda_2\kappa^2-
\frac {39}{8}\lambda_1\kappa^2+\frac 
1{4}\lambda_1^2\kappa-\lambda_1\lambda_2\kappa+4\lambda_2^3+
\frac{25}{3}\lambda_1\lambda_2^2+\frac{16}{3}\lambda_1^2\lambda_2+
\frac{22}{3}\lambda_1^3\,\,,\nonumber\\[3mm]
\beta_{\lambda_2}&=-\frac{3}{8}\lambda_2\kappa^2+\lambda_2^2\kappa+
2\lambda_1\lambda_2\kappa+\frac{7}{3}\lambda_2^3+
\frac{34}{3}\lambda_1\lambda_2^2+\frac{34}{3}\lambda_1^2\lambda_2\,\,,
\nonumber\\[3mm] 
\beta_{\lambda_3}&=\frac{3}{64}\kappa^4+204\lambda_3^2-
\frac{27}{2}\lambda_3\kappa^2+(32\lambda_1^2+32\lambda_1\lambda_2^2+
20\lambda_2)\lambda_3-(2\lambda_1\lambda_2^2+\lambda_2^3)\kappa+
\nonumber\\[3mm]
&+\frac{1}{4}(6\lambda_1^2+6\lambda_1^2\lambda_2^2+\lambda_2^2)\kappa^2+
\frac{1}{32}(24\lambda_1+12\lambda_2)\kappa^3-8\lambda_1^4-16\lambda_1^3
-28\lambda_1^2\lambda_2^2\kappa^2-
\nonumber\\[3mm]
&-20\lambda_1\lambda_2^3\kappa- 5\lambda_2^4\kappa^2\,\,.\label{befu} 
\end{align} 

A few comments about the $\beta$-functions which we are using are 
now in order. In fact, in~\cite{AKK} they represent the 
$\beta$-functions of the 
renormalization group, and were evaluated by using the minimal subtraction 
(MS) scheme. The $\beta$-functions occuring in the Callan-Symanzik equation 
(which describes the breaking of dilatations) are mass-independent, while 
those of the renormalization group (which describes the invariance of the 
model under variations of the normalization point) will in general depend 
on the mass ratios. However, a major progress in the study of 
Callan-Symanzik and renormalization group equations was iniciated by E. 
Kraus~\cite{Kr}. In her work, Kraus has observed that for an asymptotic 
normalization point (where mass effects are negleted) the $\beta$-functions 
and the anomalous dimensions of the Callan-Symanzik and renormalization group 
equations are the same to all orders of perturbation theory, and 
mass-independent if a regularization scheme like MS is used.\footnote{This 
does not occur in theories with broken symmetries. In particular, the 
Callan-Symanzik equation contains the presence of $\beta$-functions which 
correspond to the renormalization of physical masses, with the consequence that 
hard breaking of dilatations depends on the normalization point also in the 
asymptotic region~\cite{Be}.}  

In practice, we usually know the $\beta$-functions only as asymptotic 
expansions in the small coupling limit. Within this framework, it is 
convenient to introduce the notation~\cite{OSZ}:
\begin{align}
\beta_{\lambda_r}(\kappa,\lambda_r)&=C_r^{(0)}\kappa^3+
C_{r,r_1}^{(0)}\kappa^2\lambda_{r_1}
+C_{r,r_1r_2}^{(0)}\kappa\lambda_{r_1}\lambda_{r_2}+C_{r,r_1r_2r_3}^{(0)}
\lambda_{r_1}\lambda_{r_2}\lambda_{r_3}+\nonumber\\[3mm]
&+\sum_{n=4}^\infty\sum_{m=0}^nC_{r,r_1\ldots r_m}^{(n-3)}
\lambda_{r_1}\ldots\lambda_{r_m}\kappa^{n-m}\,\,,\label{befu1}
\end{align} 
where $r,r_1,\ldots,r_m=1,2$, and
\begin{align}
\beta_{\lambda_3}(\kappa,\lambda_r)&=C_3^{(0)}\kappa^4+
C_{3,r_1}^{(0)}\kappa^3\lambda_{r_1}
+C_{3,r_1r_2}^{(0)}\kappa^2\lambda_{r_1}\lambda_{r_2}+C_{3,r_1r_2r_3}^{(0)}
\kappa\lambda_{r_1}\lambda_{r_2}\lambda_{r_3}+\nonumber\\[3mm]
&+C_{3,r_1r_2r_3r_4}^{(0)}\lambda_{r_1}\lambda_{r_2}\lambda_{r_3}\lambda_{r_4}
+\sum_{n=5}^\infty\sum_{m=0}^nC_{3,r_1\ldots r_m}^{(n-4)}
\lambda_{r_1}\ldots\lambda_{r_m}\kappa^{n-m}\,\,,\label{befu2}
\end{align} 
where $r_1,\ldots,r_m=1,2,3$.

Now, replacing the expansions (\ref{befu1}), (\ref{befu2}) and (\ref{sol}) 
into the reduction equations (\ref{requa}), one finds that $f^{(0)}_{r}$, 
to the lowest order, must be a solution of the equations 
\begin{align}
H_r(f^{(0)})=C_r^{(0)}+ C_{r,r_1}^{(0)}f^{(0)}_{r_1} 
+C_{r,r_1r_2}^{(0)}f^{(0)}_{r_1}f^{(0)}_{r_2}+C_{r,r_1r_2r_3}^{(0)} 
f^{(0)}_{r_1}f^{(0)}_{r_2}f^{(0)}_{r_3}=0\,\,,\label{befu3} 
\end{align} 
where $r,r_1,\ldots,r_m=1,2$, and
\begin{align}
H_3(f^{(0)})=&C_3^{(0)}+ C_{3,r_1}^{(0)}f^{(0)}_{r_1} 
+C_{3,r_1r_2}^{(0)}f^{(0)}_{r_1}f^{(0)}_{r_2}+C_{3,r_1r_2r_3}^{(0)} 
f^{(0)}_{r_1}f^{(0)}_{r_2}f^{(0)}_{r_3}+ \nonumber\\[3mm]
&+C_{3,r_1r_2r_3r_4}^{(0)}f^{(0)}_{r_1}f^{(0)}_{r_2}f^{(0)}_{r_3} 
f^{(0)}_{r_4}=0\,\,,\label{befu4} 
\end{align} 
where $r_1,\ldots,r_m=1,2,3$.

Since we wish the Lagrangian to be bounded from below, we take into 
account only sets of real solutions which guarantee that the 
coefficient of the $(\varphi_i^*\varphi_i)^3$-term be positive. 
Therefore, the reduction 
equations (\ref{requa}) have the solutions: 
\begin{align*}
&({\rm I}) \qquad f^{(0)}_{r^\prime_1}=-0,82524\quad 
f^{(0)}_{r^\prime_2}=0,91585 
\quad f^{(0)}_{r^\prime_3}=0,092398\,\,,\\[3mm]
&({\rm II}) \qquad f^{(0)}_{r^{\prime\prime}_1}=0,09062\quad 
f^{(0)}_{r^{\prime\prime}_2}=-0,91585
\quad f^{(0)}_{r^{\prime\prime}_3}=0,092398\,\,.
\label{befu5}
\end{align*} 

Given a solution $f^{(0)}_r$ of the equations (\ref{befu3}) and 
(\ref{befu4}), we obtain for the expansion coefficients $\chi^{(m)}_r$ the 
relations
\begin{equation}
M_{rr^\prime}(f^{(0)})\chi^{(m)}_{r^\prime}=\vartheta_r\,\,.
\end{equation} 
$\vartheta_r$ depends only on the coefficients $\chi^{(p)}$, with $p \leq 
m-1$ and on the $\beta$-function coefficients, evaluated em $f^{(0)}_r$, 
for order $\leq m-1$. The matrix $M$ depends only $f^{(0)}$ and is given by 
\begin{equation}
M_{rr^\prime}(f^{(0)})=\left.\frac{\partial H_r}{\partial 
f_r}\right|_{f_r=f_r^{(0)}}\,\,. 
\end{equation} 

The lowest-order criterium to insure that all coefficients $\chi^{(m)}$ 
in (\ref{sol}) are fixed is given by:
\begin{equation}
{\rm det}\,M_{rr^\prime}(f^{(0)})\not= 0\,\,.
\end{equation} 

With the coefficients $\chi^{(m)}$ fixed, one can use, then, the reparametrization 
invariance of the theory in order to bring the $\lambda_1, \lambda_2, \lambda_3$ 
couplings into a simple form:
\begin{equation}
\lambda_r=\kappa f_r^{(0)} \quad r=1,2,3\,\,.
\end{equation}

Our matrices are:
\begin{equation*}
M_{(\mathrm{I})}=
\begin{bmatrix}
17,153 & 8,2733 & 0 \\ 
-22,034 & 7,3282 & 0 \\ 
-4,0109 & -1,6834 & -8,2072
\end{bmatrix}
\,\,,\qquad \mathrm{det}\,M_{(\mathrm{I})}\not=0\,\,,
\end{equation*}
and 
\begin{equation*}
M_{(\mathrm{II})}=
\begin{bmatrix}
16,779 & 7,3574 & 0 \\ 
-6,4852 & 3,6648 & 0 \\ 
-6,7584 & -2,5993 & -8,2072
\end{bmatrix}
\,\,,\qquad \mathrm{det}\,M_{(\mathrm{II})}\not=0\,\,.
\end{equation*}
Hence, both solutions are uniquely determined to all orders.

To end up, we would like to emphasize that our result could indicate to 
even stronger phenomenon: the possibility of getting the asymptotical 
conforme invariance. The correspondent effect in Abelian 
Chern-Simons-matter model has been discovered in~\cite{Odin}. We hope to 
report our conclusions on this in a short time.

\section{Conclusions}
 
To summarize, we conclude that the general method of reduction in the 
number of coupling parameters, when applied to the non-Abelian 
Chern-Simons-matter model with several independent couplings as considered 
here, allows us (taking into account the asymptotic region) to show that all 
$\beta$-functions vanish to any order of the perturbative series. Therefore, 
the model is asymptotically scale invariant. In forthcoming papers, we 
intend to extend this analysis to other types of topological theories, 
e.g. the BF model~\cite{bbr-pr91, LPS}, or its extension, recently 
proposed, the BFK model~\cite{Wald} coupled with matter, too. Our aim is to 
obtain results valid to all orders of perturbation theory and to classify 
the theories which can be made finite by means of the technique of 
reduction of couplings. 

\section*{Acknowledgements}

We would like to thank O. Piguet and J.A. Helay\"el-Neto for helpful 
discussions and illuminating comments. We are also grateful to the staff of 
the Group of Theoretical Physics at UCP for the warm hospitality. J.L.A. is  
grateful to staff of the DME-PUC.



\begin{thebibliography}{99}

\bibitem{bbr-pr91} D. Birmingham, M. Blau, M. Rakowski and G. Thompson, 
{\sf Phys. Rep.~}209 (1991) 129;

\bibitem{s-btc} S. Deser, R. Jackiw and S. Templeton, {\sf Ann. of 
Phys.~}140 (1982) 372;\\ A.S. Schwarz, Baku International Topological 
Conference, Abstracts, vol.2, p.345 (1987);\\ E. Witten, {\sf Commun. Math. 
Phys.~}117 (1988) 353, {\sf Commun. Math. Phys.~}118 (1988) 601, {\sf 
Commun. Math. Phys.~}121 (1989) 351; 

\bibitem{witten-grav} E. Witten, {\sf Nucl. Phys.~}B311
(1988)46; {\sf Phys. Lett.~}B206 (1988) 601;\\ S. Deser, J. McCarthy and Z. 
Yang, {\sf Phys. Lett.~}B222 (1989) 61; 

\bibitem{guadagnini}  E. Guadagnini, M. Martellini and M. Mintchev, 
{\sf Phys. Lett.~}B227 (1989) 111, {\sf Nucl. Phys.~}B330 (1990) 575; 

\bibitem{blasi}  A. Blasi and R. Collina, {\sf Nucl. Phys.~}B345
(1990) 472;\\ F. Delduc, C. Lucchesi, O. Piguet and S.P. Sorella, {\sf Nucl. 
Phys.~}B346 (1990) 313; \\C. Lucchesi and O. Piguet, {\sf Nucl. Phys.~}B381 
(1992) 281;

\bibitem{CFNP0} O.M. Del Cima, D.H.T. Franco, J.A. Helay\"{e}l-Neto 
and O. Piguet, {\sf Lett.Math.Phys.~}47 (1999) 265; 
 
\bibitem{maggiore}  A. Blasi, N. Maggiore and S.P. Sorella, 
{\sf Phys. Lett.~}B285 (1992) 54;

\bibitem{pronin}  G.A.N. Kapustin and P.I. Pronin, {\sf Phys. Lett.~}
B318 (1993) 465;

\bibitem{CFNP} O.M. Del Cima, D.H.T. Franco, J.A. Helay\"{e}l-Neto 
and O. Piguet, {\sf JHEP~}02 (1998) 002; 

\bibitem{pisor}  O. Piguet and S.P. Sorella, ``{\em Algebraic
Renormalization,}'' Lecture Notes in Physics, m28, Springer-Verlag, Berlin,
Heidelberg, 1995;

\bibitem{Bar} G. Barnich, {\sf JHEP~}12 (1998) 003; 

\bibitem{CSW} W. Chen, G.W. Semenoff and Y.S. Wu, {\sf Phys. Rev.~}D44 (1991) 
1625;

\bibitem{OSZ} R. Oehme, K. Sibold and W. Zimmermann, 
{\sf Phys. Lett.~}B147 (1984) 115;\\ W. Zimmermann, {\sf Commun. Math. 
Phys.~}97 (1985) 211;\\ R. Oehme, K. Sibold and W. Zimmermann, {\sf Phys. 
Lett.~}B153 (1985) 142;\\ R. Oehme and W. Zimmermann, {\sf Commun. Math. 
Phys.~}97 (1985) 569; 

\bibitem{Oe} R. Oehme, {\sf Prog. Theor. Phys. Suppl.~}86 (1986) 215;

\bibitem{Pi} O. Piguet, ``{\em Supersymmetry, Ultraviolet
Finiteness and Grand Unification,}'' \\ hep-th/9606045;

\bibitem{bl-pig-sor}  A. Blasi, O. Piguet and S. P. Sorella, {\sf Nucl.
Phys.~}B356 (1991) 154;

\bibitem{AKK} L.V. Avdeev, D.I. Kazakov, I.N. Kondrashuk, {\sf Nucl. 
Phys.~}B391 (1993) 333;

\bibitem{Kr} E. Kraus, {\sf Helv. Phys. Acta~}67 (1994) 425;

\bibitem{Be} C. Becchi, {\sf Commun.Math.Phys.~}47 (1973) 97;\\ 
E. Kraus, {\sf Z.Phys. C~}75 (1993) 741;\\ N. Maggiore, O. Piguet and S. 
Wolf, {\sf Nucl.Phys.~}B476 (1996) 329;\\ D.H.T. Franco, {\sf 
Phys.Rev.~}D59 (1999) 125017; 

\bibitem{Odin} S. Odintsov, {\sf Z.Phys. C~}54 (1992) 527;

\bibitem{LPS} C. Lucchesi, O. Piguet and S.P. Sorella, {\sf Nucl. 
Phys.~}B395 (1993) 325;\\M. Henneaux, {\sf Phys. Lett.~}B406 (1997) 66; 
\\A. Accardi, A. Belli, M. Martellini and M. Zeni {\sf Nucl. Phys.~}B505 
(1997) 540;\\O.M. Del Cima, D.H.T. Franco, J.A. Helay\"el-Neto and O. 
Piguet, {\sf JHEP~}04 (1998) 010;\\ R. Leitgeb, M. Schweda and H. Zerrouki, 
{\sf Nucl. Phys.~}B542 (1999) 425; 

\bibitem{Wald} O.M. Del Cima, J.M. Grimstrup and M. Schweda, {\sf Phys. 
Lett.~}B463 (1999) 48;\\ T. Pisar, J. Rant and H. Zerrouki, {\sf Mod. Phys. 
Lett.~}A15 (2000) 1147. 

\end{thebibliography}
\end{document}